\documentclass[epj,twocolumn]{webofc}
\usepackage[varg]{txfonts}   
\woctitle{LHCP 2013}

\usepackage{hyperref}
\usepackage[font=small,labelfont=bf,labelsep=period,tableposition=top]{caption}
\usepackage{mathrsfs}
\newcommand{\mh}{m_H}
\newcommand{\mt}{m_t}
\newcommand{\mz}{m_Z}

\newcommand{\sss}{\scriptscriptstyle\rm}
\renewcommand{\Re}{\mathrm{Re}\:}

\newcommand{\Lum}{\mathscr{L}}

\newcommand{\as}{\alpha_s}
\newcommand{\muf}{\mu_{\sss F}}
\newcommand{\mur}{\mu_{\sss R}}

\let\originalleft\left
\let\originalright\right
\renewcommand{\left}{\mathopen{}\mathclose\bgroup\originalleft}
\renewcommand{\right}{\aftergroup\egroup\originalright}

\newcommand{\Csoft}{C_{\text{soft}}}

\newcommand{\Chard}{C_{\text{h.e.}}}

\newcommand{\Cabf} {C_{\text{ABF}}}
\newcommand{\Csub} {C_{\text{ABF-sub}}}

\def\beq{\begin{equation}}  
\def\eeq{\end{equation}}
\def\({\left(}
\def\){\right)}
\def\[{\left[}
\def\]{\right]}

\begin{document}

\vspace*{-10ex}
\quad\\
\vspace*{-10ex}
\begin{flushright}
DESY 13-116
\end{flushright}

%
\title{An approximate N$^3$LO cross section for Higgs production in gluon fusion}
%
%

\author{
  Marco Bonvini\inst{1}
}

\institute{
Deutsches Elektronen-Synchroton, DESY, Notkestra{\ss}e 85, D-22603 Hamburg, Germany
}

\abstract{%
  An approximate expression for the inclusive Higgs production cross
  section in gluon fusion at N$^3$LO in QCD with finite top mass is
  presented.  We argue that an accurate approximation can be
  constructed combining (and improving) the large- and small-$z$
  behaviours of the partonic cross section, which are both known to
  all orders from soft-gluon (Sudakov) and high-energy (BFKL)
  resummations, respectively.  For a $125$~GeV Higgs at LHC at
  $8$~TeV, we find an increase of about $6$--$13\%$ 
  with respect to the NNLO inclusive cross section for the conventional scale
  $\mur=\mh/2$, suggesting that higher order QCD corrections might be
  underestimated by presently available results.  We also find a
  significant reduction of the scale uncertainty.  }
\maketitle
\section{Introduction}
\label{sec:intro}


In the aftermath of the discovery of the Higgs boson,
the measurement of its properties is one of the main tasks of the LHC experiments.
Accurate theoretical predictions play a fundamental role in such measurements.
However, the cross section for Higgs production through gluon fusion
(the dominant production channel) is affected by a very bad
perturbative behavior. Indeed, QCD corrections to the process, known at NLO
\cite{spira1,dawson,spira2} and NNLO
\cite{harlanderNNLO,anastasiouNNLO, ravindranNNLO, Higgsfinite, harlander1, harlander3, pak1},
give rise to very large $K$-factors,
and only a mild reduction of the scale dependence.
Therefore, the knowledge of the impact of yet higher order corrections is mandatory.

Although the computation of the N$^3$LO correction (as a soft expansion and in the large $\mt$ limit)
to the cross section is in progress~\cite{anastasiouintegrals,Hoschele:2012xc,Anastasiou:2013srw,Buehler:2013fha},
in Ref.~\cite{Ball:2013bra} we have derived an approximate expression for it.
Our approximation is based on combining a soft approximation and a high-energy approximation,
taking into account the exact $\mt$ dependence.
The consistency of the combination gives some constraints
which turn out to improve significantly the accuracy
and the reliability of the result.

In the following we briefly review the basic properties of our approximation,
while we refer the Reader to Ref.~\cite{Ball:2013bra} for a detailed discussion,
and we present some new unpublished results, namely we show our prediction
for higher LHC energies and we discuss the impact of the virtually largest unknown
contribution at order $\as^3$.

\section{Combined soft and high-energy approximations}
\label{sec:approx}

The inclusive Higgs production cross section is given as a sum over partons of
convolutions of a parton luminosity $\Lum_{ij}(x,\muf^2)$ and a coefficient function $C_{ij}$,
\begin{multline}
\sigma(\tau,\mh^2) = \tau\, \sigma_0(\mh^2,\as)\sum_{i,j=\{q,g\}}\int_\tau^1
\frac{dz}z \, \Lum_{ij}\(\frac{\tau}{z},\muf^2\)\\
\times C_{ij}\(z,\as,\mur^2/\mh^2,\muf^2/\mh^2\),
\end{multline}
where $\as\equiv\as(\mur^2)$, $\sigma_0(\mh^2,\as)$ is the leading order partonic cross section,
$\tau=\mh^2/s$ and $\muf$, $\mur$ are the factorization and renormalization scales.
The sum over partons is dominated by $i=j=g$, the other channels giving a contribution
of about $3\%$ of the total result at NNLO.
Concentrating therefore on the $gg$ channel, and suppressing the subscripts and the
dependence on the scales, we have the expansion
\beq
C(z,\as) = \delta(1-z) + \as C^{(1)}(z) + \as^2 C^{(2)}(z) + \as^3 C^{(3)}(z) + \ldots
\eeq
Currently, the NLO coefficient $C^{(1)}(z)$ is known exactly~\cite{spira2},
while the NNLO coefficient $C^{(2)}(z)$ is known either in the large-$\mt$
effective theory~\cite{harlanderNNLO,anastasiouNNLO, ravindranNNLO} or
as an expansion in powers of $\mh/\mt$~\cite{Higgsfinite, harlander1, harlander3, pak1}.

In Ref.~\cite{Ball:2013bra} we have constructed an approximation
to the N$^3$LO coefficient based on the decomposition
\beq\label{eq:Capprox3}
C^{(3)}_{\rm approx}(z) = \Csoft^{(3)}(z) + \Chard^{(3)}(z),
\eeq
where $\Csoft$ contains terms predicted by soft-gluon (Sudakov) resummation
and reproduces the $z\to1$ behavior,
and $\Chard$ contains terms predicted by high-energy (BFKL) resummation
and describes the $z\to0$ limit.
In $N$ space, where $N$ is the variable conjugate to $z$ by Mellin transformation,
$\Csoft^{(3)}(N)$ reproduces the logarithmic large $N$
behavior and $\Chard^{(3)}(N)$ the rightmost singularity in $N=1$.
In principle, if all the singularities in the complex $N$ plane were known,
it would be possible to reconstruct the function everywhere;
in practice, the knowledge of the \emph{dominant} singularities ($N=\infty$ and $N=1$)
is sufficient to give a good approximation of the function in the physical region
$\Re N>1$, the other singularities, placed at integer non-positive values of $N$,
giving a \emph{subdominant} contribution.

In the following we give some details on the construction of both
approximations, with an emphasis on the improvements over the commonly
used results, and we will demonstrate the reliability of the
approximations by comparison to the first two known perturbative
orders.

\begin{figure*}
\centering
\includegraphics[width=0.495\textwidth,clip]{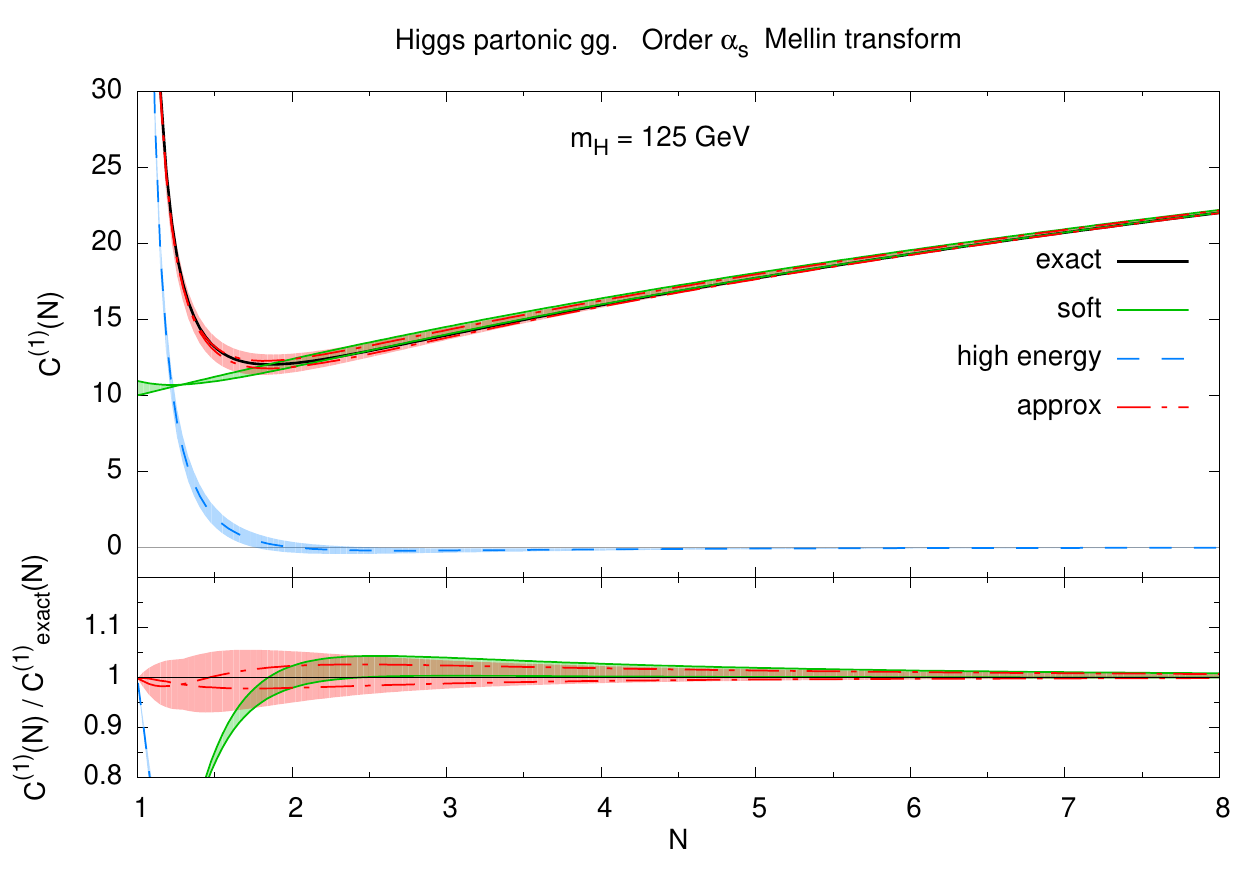}
\includegraphics[width=0.495\textwidth,clip]{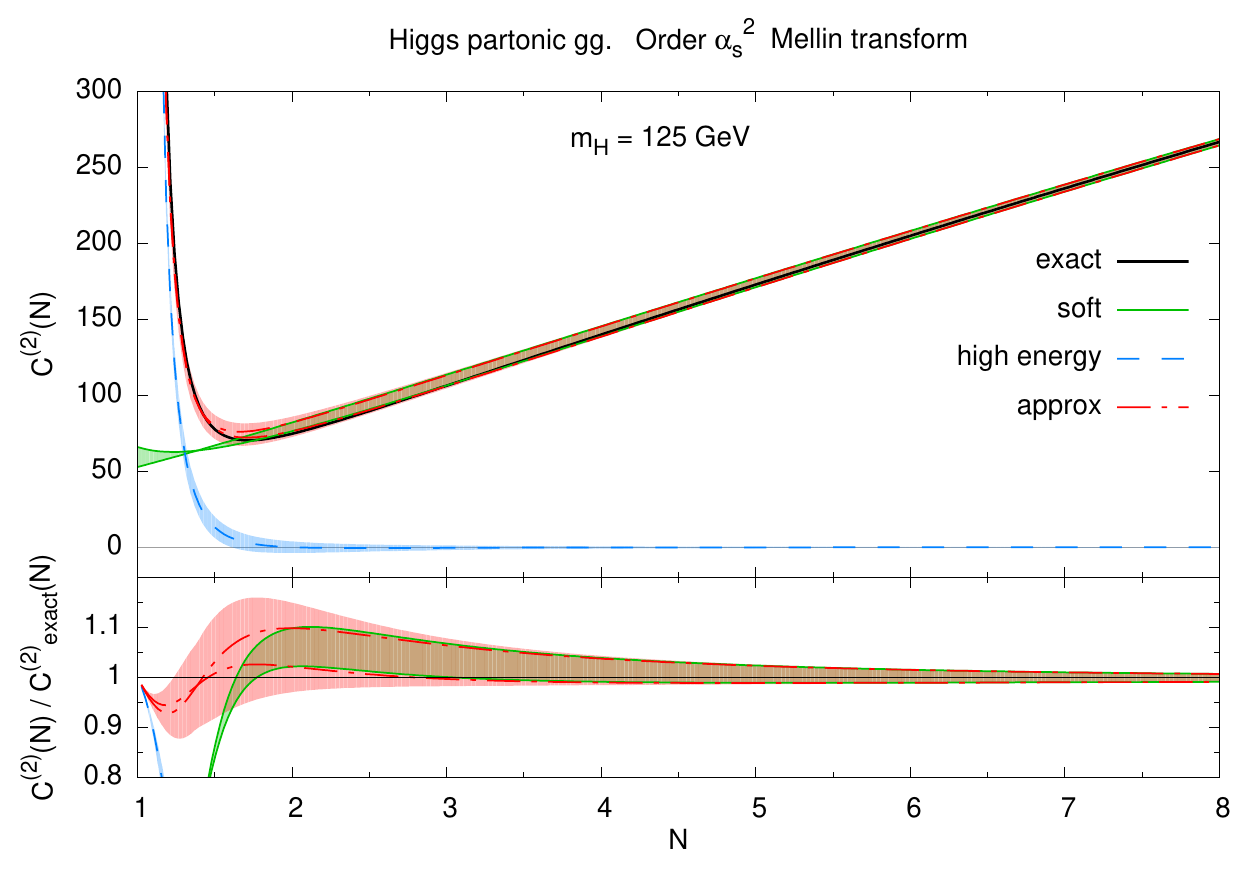}
\caption{Comparison of the NLO (left) and NNLO (right) exact
  coefficient functions to their soft, high-energy and combined approximations.}
\label{fig:partonic}
\end{figure*}

\subsection{Improved soft approximation}
\label{sec:soft}

The largest contribution at the current and foreseeable future collider energies
comes from the soft part. This is mainly due to the fact that the gluon-gluon
luminosity, peaked at small $x$, enhances the contribution from the soft region
of the partonic coefficient. This statement can be made quantitative in $N$ space,
by means of a saddle point argument~\cite{bfrsaddle}, noticing that the
region of $N$ contributing to the hadron-level cross section is localized
about the position of the saddle point of the Mellin inversion integrand.
For $\mh=125$~GeV and $\sqrt{s}=8$~TeV, the saddle point is at $N\sim2$,
a value at which the high-energy contribution is small (see Sect.~\ref{sec:he}).

On the other hand, the region $N\sim2$ cannot be regarded as a large $N$ region,
given that terms suppressed by one or more powers of $1/N$
with respect to the logarithmically growing soft terms, hereafter called subdominant,
usually negligible
at large $N$, give in general a sizable contribution at $N\sim2$.
Therefore, the quality of a soft approximation at the saddle point
strictly depends on the control on such subdominant terms.
In fact, some of these terms are under control to all orders in perturbation theory~\cite{Kramer:1996iq},
and their inclusion in a soft approximation leads to a significant improvement
of the quality of the approximation itself.

Our soft approximation is constructed as follows.
Starting from a resummed (all-order) soft expression in $N$ space,
we expand it in powers of $\as$, obtaining a linear combination of powers
of $\log N$. We then take the inverse Mellin transform, which is a linear combination
of logarithms of the form\footnote{We omit the details related to the distributional nature of the logarithms,
for which we refer to the original work~\cite{Ball:2013bra}.}
\beq\label{eq:wrong_log}
\frac{\log^k\log\frac1z}{\log\frac1z}
\eeq
for integer values of $k$.
We then apply the following two improvements:
\begin{itemize}
\item We use the exact logarithmic terms as originating from the kinematics of gluon emissions,
which have the form
\beq
\frac{\log^k\frac{1-z}{\sqrt{z}}}{1-z}.
\eeq
Usually the factor $\sqrt{z}$ is neglected in the soft limit,
and in resummed expressions the Mellin transform of such terms,
which is given by polygamma functions, is approximated with powers of $\log N$.
This second approximation, in particular, is incompatible with the analytic
structure of the coefficient function, since $\log N$ has a branch-cut
for negative real $N$ while polygamma functions have just poles in integer
non-positive values of $N$.
\item We include for each emission the appropriate (leading) Altarelli-Parisi
splitting kernel,\footnote{Since in Sudakov resummation the single emission
exponentiates, this factor is in fact included at the exponent, before expanding
in powers of $\as$.}
\beq
p_{gg}(z) = \frac{A_g(z)}{1-z}, \quad
A_g(z) = \frac{C_A}\pi \frac{1-2z+3z^2-2z^3+z^4}{z}.
\eeq
Usually, $A_g(z)$ is approximated with $A_g(1)$, and only the divergent part of
$p_{gg}$, $(1-z)^{-1}$, is kept.
However, we point out (following an observation in Ref.~\cite{Kramer:1996iq})
that the inclusion of subleading terms in $A_g(z)$ significantly improves the soft
approximation.
In fact, the exact expression of $A_g(z)$ would introduce a double counting
with high-energy terms (because of the $z^{-1}$ term), and we therefore
use an expansion of $A_g(z)$ about $z=1$ to order $1$ and $2$
(above $2$ there is no practical difference in the region $N\gtrsim2$).
The difference between these two results gives information on the size
of the subdominant terms that are not included, and therefore provides a measure
of the uncertainty associated with the soft approximation.
\end{itemize}
In Fig.~\ref{fig:partonic} the NLO and NNLO coefficient functions
$C^{(1)}(N)$ and $C^{(2)}(N)$ are shown together with their soft approximations.
The two green curves (filled with a green band in between)
represent the construction of the soft approximation described above,
for the expansion of $A_g(z)$ to first and second order in $(1-z)$.
It is clear from the plots that the soft approximation reproduces well
the exact coefficient for $N\gtrsim2$, the approximation obtained
expanding to second order $A_g(z)$ (the lower green curve at large $N$)
being closer to the exact result.

\subsection{High-energy approximation}
\label{sec:he}

The leading pole in $N=1$ at each power of $\as$ is predicted
to all orders by BFKL resummation~\cite{Ball:2007ra,abfquarks},
and can be obtained from the all-order formula
\beq\label{eq:ABF}
\Cabf(N,\as) = \sum_{\substack{i_1,i_2 \geq 0\\i_1+i_2>0}} c_{i_1,i_2} \[\gamma_+^{i_1}(N,\as)\] \[\gamma_+^{i_2}(N,\as)\],
\eeq
where $\gamma_+$ is the largest eigenvalue of the DGLAP singlet anomalous dimension matrix,
the square brackets symbolize the inclusion of running coupling effects~\cite{Ball:2007ra,abfquarks},
and $c_{i_1,i_2}$ are numeric coefficients that have been computed to the first
few orders in Refs.~\cite{Higgsfinite,marzaniPhD}.
Eq.~\eqref{eq:ABF} predicts correctly the coefficient of the highest order $N=1$ pole
at each power of $\as$, provided the anomalous dimension is accurate at the same
(leading logarithmic) level. However, for a consistent resummed result
it is more convenient to use the resummed anomalous dimension, because the
resummation changes the position of the leading pole.
Since the resummed anomalous dimension vanishes in $N=2$ (momentum conservation)
this implies, in turn, that $\Cabf(2,\as)$ would vanish as well.

For our purpose, i.e.\ obtaining a high-energy approximation to the coefficient
function at a finite perturbative order, we consider the expansion of Eq.~\eqref{eq:ABF}:
\beq\label{eq:ABFexp}
\Cabf(N,\as) = \sum_{n=1}^\infty \as^n \Cabf^{(n)}(N).
\eeq
The computation of $\Cabf^{(3)}(N)$ requires the anomalous dimension $\gamma_+$
up to order $\as^3$. We could use the exact anomalous dimension, which is known;
however, it grows logarithmically at large $N$,
and this would interfere with the soft approximation, spoiling its accuracy.
Therefore we adopt the following procedure:
\begin{itemize}
\item We take an expansion of the anomalous dimension about $N=1$
to NLL order (namely the largest and the next-to-largest
pole at each order in $\as$).
We stress that the final result, $\Cabf^{(n)}(N)$, is still accurate at LL only,
though the NLL terms included in this way may (and do) improve the accuracy of the approximation.
\item Since $\Cabf^{(n)}(N)$ constructed in this way still doesn't vanish at large $N$,
we subtract the large $N$ terms,
\beq \label{eq:Charddamp}
\Csub^{(n)}(N)= \Cabf^{(n)}(N)-2 \Cabf^{(n)}(N+1)+ \Cabf^{(n)}(N+2),
\eeq
introducing spurious poles at integers $N\leq0$, hereafter called subdominant, which are
beyond our control. This subtraction corresponds to a $z$-space damping $(1-z)^2$.
\item Finally, since the momentum conservation property of $\Cabf$ is lost in this procedure,
we restore it by hand adding a subdominant term,
\begin{align} \label{eq:Chardmom}
&\Chard^{(n)}(N)= \Csub^{(n)}(N)- \frac{4 !\, k_{\rm mom}}{N (N+1)(N+2)},
\end{align}
where $k_{\rm mom}$ must equal $\Csub^{(n)}(2)$.
In order to estimate the impact of subdominant poles, we assign an arbitrary $5\%$
uncertainty to $k_{\rm mom}$, using $\Csoft(2)$ as a reference:
\beq\label{momconst}
k_{\rm mom}=\Csub(2) \pm 0.05\times \Csoft(2).
\eeq
\end{itemize}
The results at NLO and NNLO are shown in Fig.~\ref{fig:partonic}.
The high-energy part alone is accurate only very close to the singularity in $N=1$,
while it vanishes fast for $N\gtrsim2$.
The combination $\Csoft+\Chard$, red curves (corresponding to the two soft curves),
is instead very accurate in the whole range $N>1$.
The red band is the (linear) combination of the soft and high-energy uncertainties,
and represents our final estimate of the error from neglected subdominant terms.


We conclude that our construction is robust and gives
accurate approximations to the first two orders in perturbation theory.
Arguing that this feature remains true at higher orders, we have constructed
an approximate expression for the third order coefficient, Eq.~\eqref{eq:Capprox3}.
In the next Section we will use it to predict the N$^3$LO cross section.

\begin{figure*}[t]
\centering
\includegraphics[width=0.495\textwidth,clip]{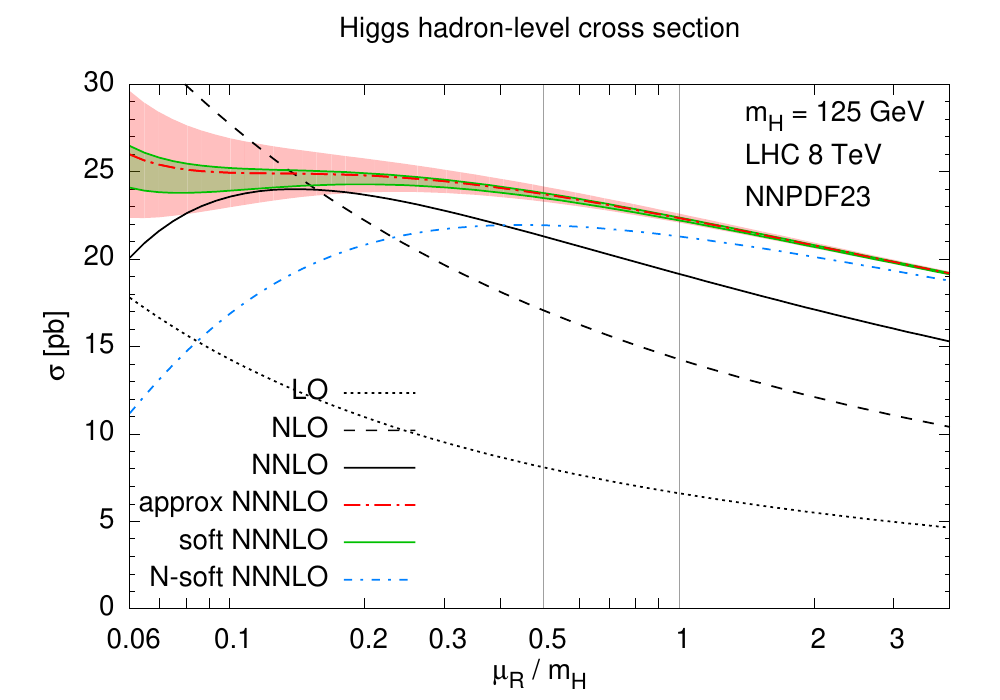}
\includegraphics[width=0.495\textwidth,clip]{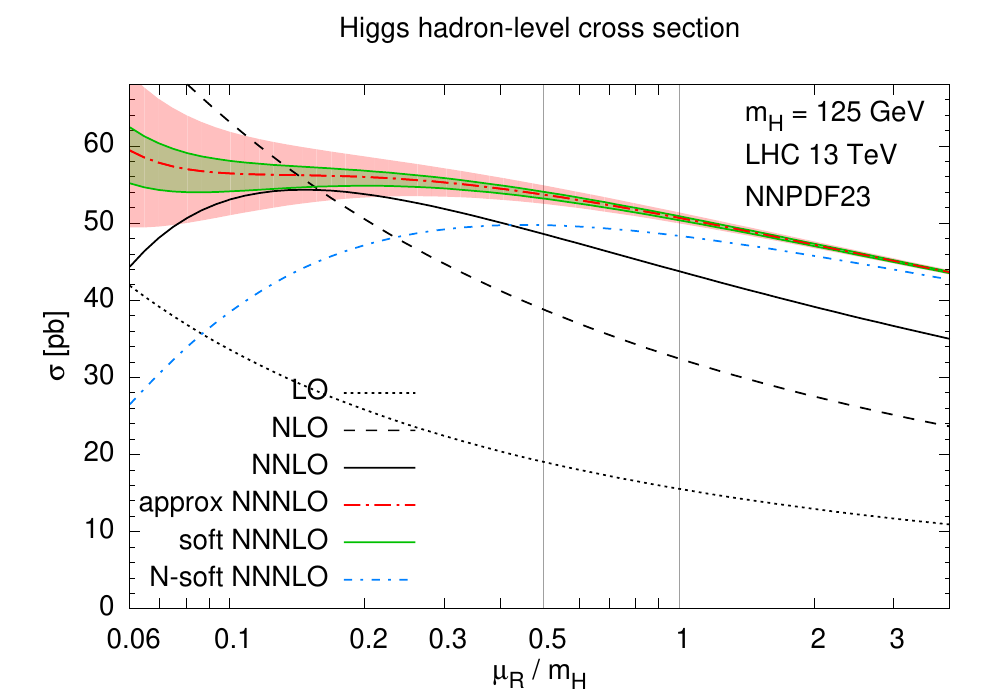}
\includegraphics[width=0.495\textwidth,clip]{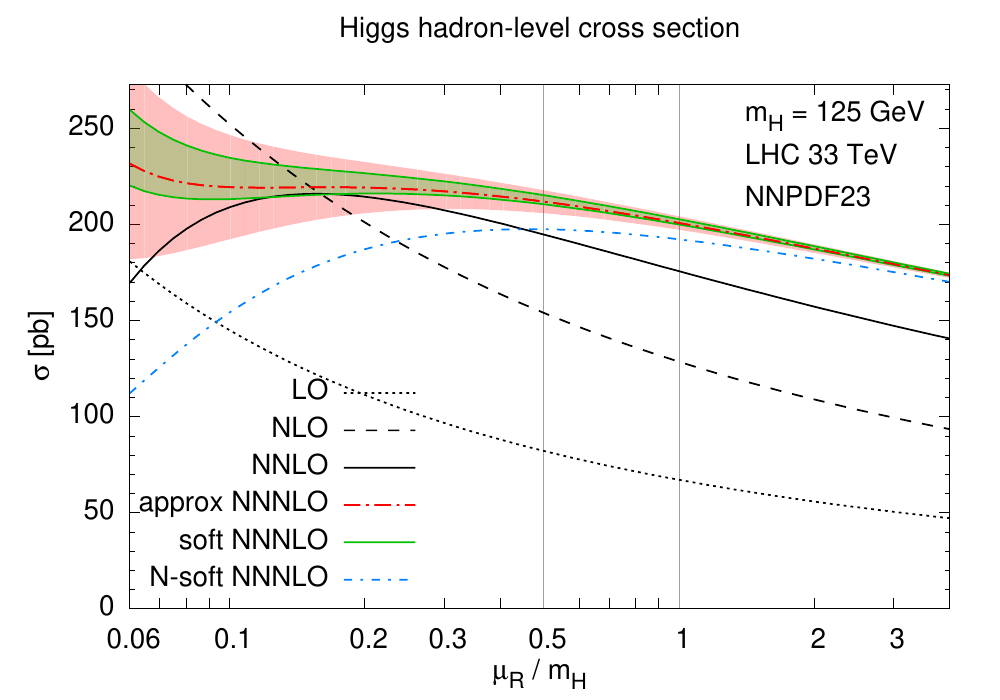}
\includegraphics[width=0.495\textwidth,clip]{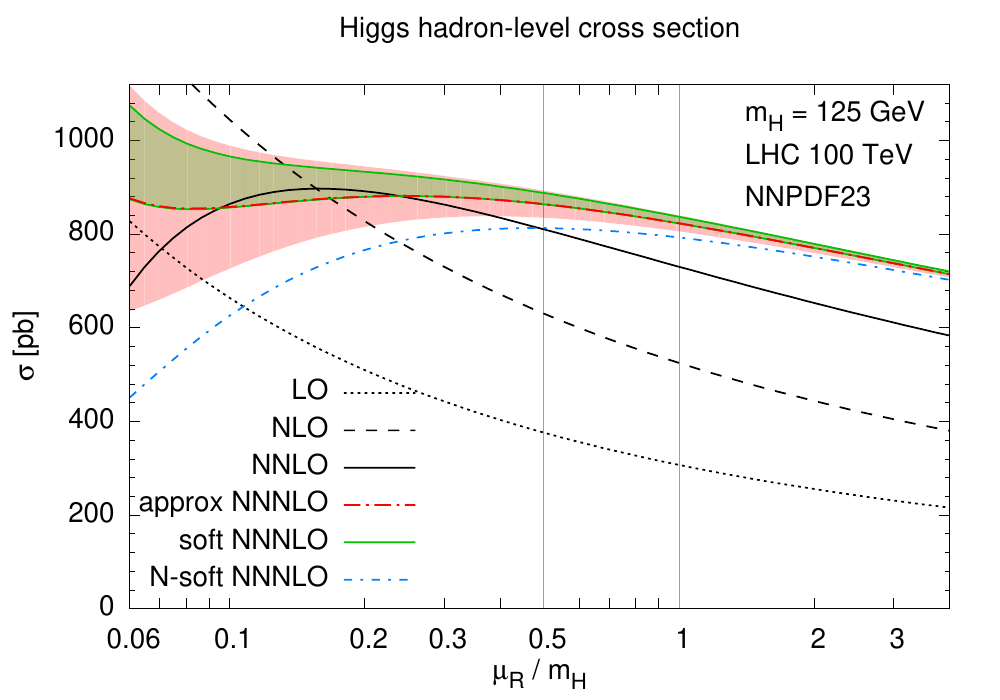}
\caption{Dependence of the N$^3$LO cross section on the renormalization scale~$\mur$,
  for $\muf=\mh=125$~GeV and collider energy $8$~TeV (top-left), $13$~TeV (top-right),
  $33$~TeV (bottom-left) and $100$~TeV (bottom-right).
  The two standard choices of renormalization scale are shown as vertical lines.}
\label{fig:ren}
\end{figure*}

\section{Results}
\label{sec:results}

We present here the result for the production cross section
of a Higgs boson with mass $\mh=125$~GeV at the LHC for several collider energies.
Differently from Ref.~\cite{Ball:2013bra}, we use here the
NNPDF2.3 pdf set~\cite{Ball:2012cx}, with $\as(\mz^2)=0.118$.
Results are obtained with the code \texttt{ggHiggs}~\cite{code}.

In Fig.~\ref{fig:ren} we show the dependence of the cross section
on the renormalization scale $\mur$, keeping the factorization scale
fixed $\muf=\mh$ (in this way we capture the largest dependence at low collider energies,
since the factorization scale dependence is very mild over a wide
range~\cite{Ball:2013bra,Buehler:2013fha}).
In addition to the exact (with exact $\mt$ dependence) LO, NLO and NNLO,
we show our approximation for the N$^3$LO cross section (red curve)
with the estimated error from the unknown subdominant terms (red band)
as described in Sect.~\ref{sec:approx}.
We observe that our prediction corresponds to an increase of the cross section
which ranges from about $17\%$ to about $13\%$ for collider energies
from $8$~TeV to $100$~TeV at the central scale $\mur=\mh$,
while the increase is lower at the scale $\mur=\mh/2$, from
about $11\%$ to about $7\%$ for the same energies.

To show the impact of the soft and high-energy terms separately,
we also plot the approximation obtained considering the soft terms only
(green lines and band).
For instance, at $\sqrt{s}=13$~TeV (next LHC energy) the impact in our central prediction
of the high-energy terms is minimal, since the red curve lies almost
exactly in the middle of the soft green band.
At higher collider energies, the high-energy terms become more relevant,
consistently to the fact that at higher energies the saddle point
moves to lower values~\cite{bfrsaddle} (about $N\sim1.7$ at $\sqrt{s}=100$~TeV).

The plots show another curve denoted $N$-soft, which corresponds to a soft
approximation as obtained if we do not apply the two improvements described
in Sect.~\ref{sec:soft}. Such a curve is conceptually identical to the result
of Ref.~\cite{MV}, except some minor details.
Therefore, the difference between our soft curves and the $N$-soft curve
is entirely due to the improvements we have introduced.
Such improvements turn out to predict a larger cross section,
and give a flatter scale dependence in the region of low $\mur$,
where some sort of convergence of the perturbative expansion shows up.

We finally discuss the impact of unknown \emph{dominant}
terms. Indeed, the error band in our result accounts for \emph{subdominant}
(i.e., not fixed by the resummation formalism)
unknown contributions, but in fact some dominant terms are unknown.
As discussed in Sect.~\ref{sec:he}, the high-energy approximation is accurate
only at LL level, but at order $\as^3$ also NLL and NNLL contribute.
These contributions are definitely important close to $N=1$,
but they are likely of the same order of subdominant terms
at $N\sim2$ (where the saddle point lies for $\sqrt{s}=8$~TeV),
and are therefore already taken into account by our uncertainty band.
On the other hand, at order $\as^3$ all the soft logarithmic terms are known,
but the $\delta(1-z)$ coefficient, corresponding in $N$ space to a constant term, is unknown.
\begin{figure}
\centering
\includegraphics[width=0.495\textwidth,clip,page=3]{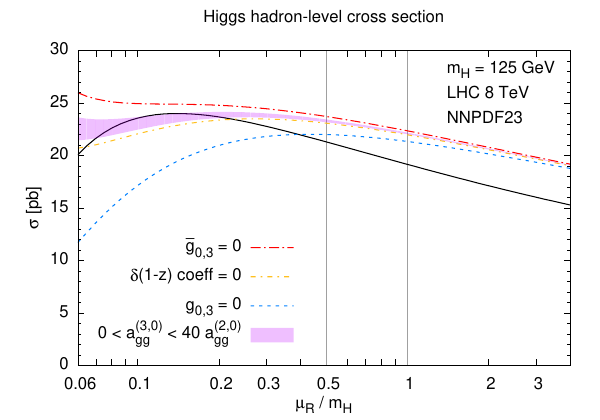}
\caption{Same as Fig.~\ref{fig:ren} for $\sqrt{s}=8$~TeV, using different guesses
  for the unknown $\delta(1-z)$ term at order $\as^3$.}
\label{fig:g03}
\end{figure}
To investigate the impact of such unknown term,
in Fig.~\ref{fig:g03} we show several possibilities:
\begin{itemize}
\item In our approximation we have kept as a default choice all the constant terms in $N$
  space coming from the Mellin transform of the plus distributions,
  and we have set to zero all the others.
  According to the terminology introduced in Ref.~\cite{Ball:2013bra},
  we have set $\bar g_{0,3}=0$.
\item This choice is conceptually similar to omitting the
  coefficient of $\delta(1-z)$ at order $\as^3$, as suggested in Ref.~\cite{MV}.
\item A third possibility, which would be adopted in a naive NNNLL resummation,
  would be to set to zero all the constant terms in $N$-space, $g_{0,3}=0$.
\item Another possibility is to estimate the $\delta(1-z)$ coefficient
  from lower orders~\cite{Buehler:2013fha}. In the large $\mt$ effective theory,
  such coefficient factorizes into a Wilson coefficient and a ``pointlike''
  expansion in powers of $\as/\pi$, whose coefficients have been called $a_{gg}^{(n,0)}$
  in Ref.~\cite{Buehler:2013fha}. Up to order $\as^2$ such coefficients are well behaved,
  with $a_{gg}^{(1,0)}/a_{gg}^{(0,0)}\sim10$ and $a_{gg}^{(2,0)}/a_{gg}^{(1,0)}\sim1.4$, and it then appears reasonable
  to estimate $0<a_{gg}^{(3,0)}/a_{gg}^{(2,0)}<40$, as proposed in Ref.~\cite{Buehler:2013fha}.
\end{itemize}
Except the option $g_{0,3}=0$, which looks unreasonable in the light of the
large coefficients of the perturbative expansion of $g_0(\as)$~\cite{Ball:2013bra},
all the other options are quite close each other, and give an uncertainty comparable
to the one from subdominant terms.
It is interesting to observe that our default option predicts the largest cross section,
while the smaller (reasonable) prediction is obtained setting to zero the coefficient
of the $\delta(1-z)$ term, which would decrease the size of the N$^3$LO contribution
from about $11\%$ to about $8\%$ at $\mur=\mh/2$.
Anyway, the difference among these predictions can only set the size of
the uncertainty associated with the unknown coefficient,
though only the computation of such coefficient (even in the large $\mt$ effective theory)
can solve the ambiguity.

\section{Conclusions}

We have constructed an approximation to the Higgs production cross section
combining and improving soft and high-energy behaviors.
At the known orders, such approximation accurately reproduces the exact result
within the estimated error coming from subdominant terms.
We have then used it to predict the N$^3$LO cross section.
The largest uncertainty on our approximation comes from the unknown dominant term proportional
to $\delta(1-z)$, whose impact has been studied in some detail,
and whose uncertainty can only be fixed by its computation.
Taking into account all the uncertainties, we can reasonably conclude
that the N$^3$LO correction amounts to a $6$--$13\%$ increase
over the NNLO at the conventional scale $\mur=\mh/2$
for $\sqrt{s}=8$~TeV.
We also note that the scale uncertainty in the conventional range
$\mh/4<\mur<\mh$ is reduced from $4.1$~pb at NNLO to $2.3$~pb at N$^3$LO,
which is rather larger than the uncertainty on our approximation.
This proves that our result, though approximate, reduces the
uncertainty on the Higgs cross section, and provides therefore a
step forward in the Higgs precision phenomenology task.


%
%

\vspace{-9em}
\end{document}